\documentclass[conference]{IEEEtran}
\IEEEoverridecommandlockouts

\usepackage{cite}
\usepackage{booktabs}
\usepackage{float}
\usepackage{amsmath,amssymb,amsfonts}
\usepackage{algorithmic}
\usepackage{graphicx}
\usepackage{textcomp}
\usepackage{xcolor}
\usepackage{multirow}
\usepackage{hyperref} 
\usepackage{paralist}

\usepackage{diagbox}
\usepackage{footmisc} 

\def\BibTeX{{\rm B\kern-.05em{\sc i\kern-.025em b}\kern-.08em
    T\kern-.1667em\lower.7ex\hbox{E}\kern-.125emX}}
\begin{document}

\title{Adversarial Alignment and Disentanglement for Cross-Domain CTR Prediction with Domain-Encompassing Features
}

\author{
\IEEEauthorblockN{Junyou He\IEEEauthorrefmark{1}\thanks{\IEEEauthorrefmark{1} Both authors contributed equally to this research.}}
\IEEEauthorblockA{
\textit{Search and Recommendation } \\
\textit{Platform Department, JD.COM}\\
Beijing, China \\
hejunyou1@jd.com}
\and
\IEEEauthorblockN{Lixi Deng\IEEEauthorrefmark{1}\IEEEauthorrefmark{2}\thanks{\IEEEauthorrefmark{2} To whom correspondence should be addressed.}}
\IEEEauthorblockA{
\textit{Search and Recommendation } \\
\textit{Platform Department, JD.COM}\\
Beijing, China \\
denglixi@jd.com}
\and
\IEEEauthorblockN{Huichao Guo}
\IEEEauthorblockA{
\textit{Search and Recommendation } \\
\textit{Platform Department, JD.COM}\\
Beijing, China \\
guohuichao1@jd.com}
\and
\hspace{1.3cm} 
\IEEEauthorblockN{Ye Tang}

\IEEEauthorblockA{
\hspace{1.3cm} 
\textit{Petro-CyberWorks Information} \\
\hspace{1.3cm} 
\textit{Technology Co., Ltd.}\\
\hspace{1.3cm} 
Beijing, China \\
\hspace{1.3cm} 
feelingandfeeling@qq.com}
\and
\hspace{1.3cm} 
\IEEEauthorblockN{Yong Li}
\IEEEauthorblockA{
\hspace{1.3cm} 
\textit{Search and Recommendation } \\
\hspace{1.3cm} 
\textit{Platform Department, JD.COM}\\
\hspace{1.3cm} 
Beijing, China \\
\hspace{1.3cm} 
liyong5@jd.com}
\and
\hspace{1cm} 
\IEEEauthorblockN{Sulong Xu}
\IEEEauthorblockA{
\hspace{1cm} 
\textit{Search and Recommendation } \\
\hspace{1cm} 
\textit{Platform Department, JD.COM}\\
\hspace{1cm} 
Beijing, China \\
\hspace{1cm} 
xusulong@jd.com}
}

\maketitle

\begin{abstract}
Cross-domain recommndation (CDR) has been increasingly explored to address data sparsity and cold-start issues. However, recent approaches typically disentangle domain-invariant features shared between source and target domains, as well as domain-specific features for each domain. However, they often rely solely on domain-invariant features combined with target domain-specific features, which can lead to suboptimal performance. 
To overcome the limitations, this paper presents the Adversarial Alignment and Disentanglement Cross-Domain Recommendation ($A^2DCDR$ ) model, an innovative approach designed to capture a comprehensive range of cross-domain information, including both domain-invariant and valuable non-aligned features. The $A^2DCDR$ model enhances cross-domain recommendation through three key components: refining MMD with adversarial training for better generalization, employing a feature disentangler and reconstruction mechanism for intra-domain disentanglement, and introducing a novel fused representation combining domain-invariant, non-aligned features with original contextual data. Experiments on real-world datasets and online A/B testing show that $A^2DCDR$ outperforms existing methods, confirming its effectiveness and practical applicability. The code is provided at \url{https://github.com/youzi0925/A-2DCDR/tree/main}.

\end{abstract}

\section{Introduction}

\begin{figure}[htbp]
  \centering
  \includegraphics[width=\linewidth]{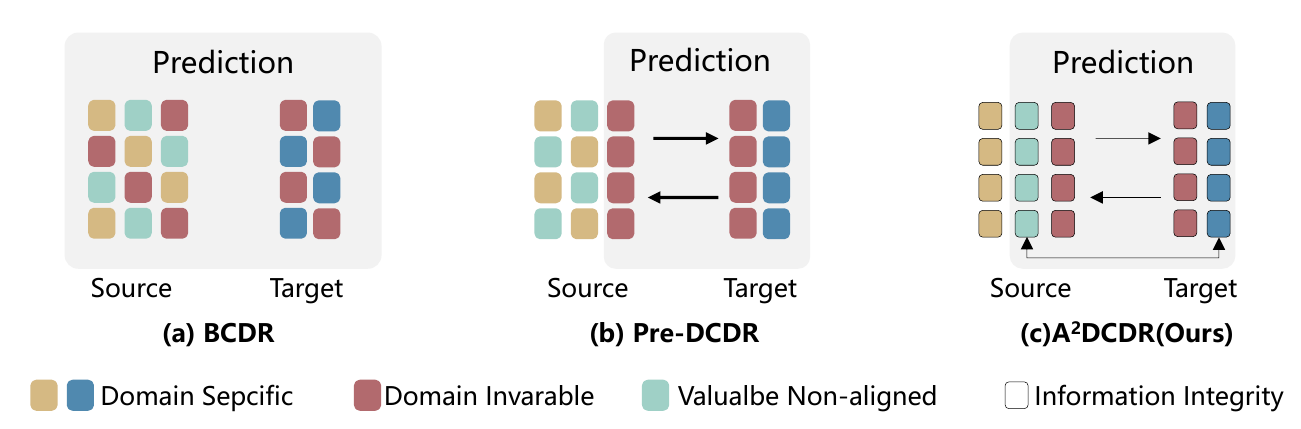} 
   \caption{Illustration of existing cross-domain recommendation (CDR) approaches: (a) Blended CDR methods, which blend all representations across domains including noise; (b) Disentangled CDR methods, which focus on separating domain-invariant and domain-specific information; and (c) The proposed $A^2DCDR$ framework, which captures both domain-invariant and valuable non-aligned features through adversarial training and feature reconstruction.}
  \label{fig:existing CDR methods}
\end{figure}


Recommender systems are critical for helping users discover relevant items in vast information spaces \cite{zang2022mpan}. However, traditional systems struggle with cold start and data sparsity issues \cite{pan2019warm, DBLP:conf/kdd/WeiLLDLW23}, which limit their effectiveness. Inspiried by the success of cross-domain learning\cite{yang2014crossmm, taigman2016unsupervisedimage, DBLP:conf/kdd/ZhangWYZ022, zhao2024all, song2024mitigating, ZhaoL0W24}, cross-domain recommendation (CDR) has emerged as a powerful solution by leveraging user preferences across domains \cite{zhu2019dtcdr, li2020ddtcdr,DBLP:conf/kdd/Xie0WLZL22, cao2022disencdr, zhang2023disentangled, DBLP:conf/www/00050LHSTW24, DBLP:conf/www/XuWWHMCHY24, DBLP:conf/www/YangJWLSWTZ24, kim2024fast, DBLP:conf/kdd/AnGYTWHZGG24, ma2024cross}. 

Cross-domain recommendation (CDR) approaches can be broadly categorized into blended and disentangled methods. Blended CDR methods, as shown in Fig. \ref{fig:existing CDR methods}(a), transfer and combine learned representations across domains. For example, DTCDR \cite{zhu2019dtcdr} enhances recommendations by sharing embeddings of common users, while DDTCDR \cite{li2020ddtcdr} uses orthogonal mapping to preserve user preferences and inter-domain similarities. GA-DTCDR \cite{2020A} further enriches embeddings via graph-based techniques and element-wise attention. However, these methods often fail to distinguish between source and target domain preferences, leading to intertwined representations and noisy information in the target domain, which degrades recommendation accuracy and interpretability.

Recent advances have shifted toward disentangling domain-invariant and domain-specific information during cross-domain knowledge transfer, as depicted in Fig. \ref{fig:existing CDR methods}(b). DisenCDR \cite{cao2022disencdr} leverages variational inference with KL divergence, while DCCDR \cite{zhang2023disentangled} employs contrastive learning. DDCDR \cite{DBLP:conf/kdd/AnGYTWHZGG24} introduces a teacher-student framework for selective information transfer. GDCCDR\cite{liu2024graph} proposes a Graph Disentangled Contrastive framework for CDR with personalized transfer by meta-networks.
Although these methods effectively align shared knowledge, their focus on distribution alignment inherently excludes non-aligned features. While many non-aligned features are inconsistent across domains, some—termed valuable non-aligned features—can still benefit the target domain. Thus, relying solely on domain-invariant information is insufficient to address nuanced needs of users.

To address these limitations,  we introduce the $\textbf{A}$dversarial  $\textbf{A}$lignment and  $\textbf{D}$isentanglement  $\textbf{C}$ross-$\textbf{D}$omain  $\textbf{R}$ecommendation ($A^2DCDR$), as illustrated in Fig. \ref{fig:existing CDR methods}(c). The black border in Fig. \ref{fig:existing CDR methods}(c) indicates information integrity.
Unlike existing methods that focus solely on domain-invariant features, $A^2DCDR$ captures a broader range of information, termed domain-encompassing feature, which includes both domain-invariant features and latent valuable non-aligned features. Non-aligned features preserve source-exclusive signal scritical for target domain predictions. For example, in the context of Sport $\rightarrow$ Cloth, the source domain (Sport) captures the signal of ‘weekly gym frequency’ (non-aligned), while the target domain (Cloth) lacks this signal but benefits from it(e.g.,gym-goers tend to prefer moisture-wicking fabrics).
Our approach uses adversarial training for cross-domain alignment and feature reconstruction for intra-domain disentanglement, retaining more effective information.
Specifically, the $A^2DCDR$ model employs inter-domain adaptation to align invariant representations and incorporates adversarial learning to enrich the source representation. By including more task-specific information\cite{Caruana98Multitask}, this module improves the generalization of the learned representations, thus preserving more valuable information from the source domain and not being limited to domain-invariant features.
Furthermore, an intra-domain disentangler with a mutual information minimizer enhances the purity of disentangled representations, while a feature reconstructor preserves latent valuable non-aligned features.
Finally, recognizing the unique value of both domain-invariant and valuable non-aligned representations,  we strategically fuse them with contextual features for the final prediction. Experimental results show that $A^2DCDR$ significantly outperforms state-of-the-art methods, setting new benchmarks in cross-domain recommendation.

In summary, our main contributions are as follows:
\begin{itemize}
\item \textbf{Comprehensive Inter-Domain Adaptaion:} $A^2DCDR$ goes beyond focusing solely on domain-invariant features by also incorporating valuable non-aligned features. This allows the model to capture a wider range of significant cross-domain information, leveraging adversarial training to improve the accuracy of recommendations.
\item  \textbf{Enhanced Intra-Domain Disentanglement:} The model introduces an intra-domain disentangler to effectively separate domain-invariant and domain-specific representations. By incorporating a mutual information minimizer, it ensures the purity of these disentangled representations, while a feature reconstructor maintains the integrity of the information during the disentanglement process.
\item \textbf{Thorough Experimental Validation:} Through extensive experiments on real-world datasets and rigorous online A/B testing, we demonstrate that $A^2DCDR$ significantly outperforms existing state-of-the-art methods. This empirical validation confirms the effectiveness of our proposed mechanisms in improving recommendation accuracy and showcases the practical applicability of our model in real-world scenarios.
\end{itemize}

\section{Related Work}
In this section, we partition the existing CDR architectures into two groups: blended CDR and disentangled CDR.
\subsection{Blended Cross-Domain Recommendation}
Blended CDR methods mainly introduce different transferring layers to transfer and blend shared knowledge across domains. For example, CoNet\cite{hu2018conet} introduces cross-connection network to conduct a dual knowledge transfer. DTCDR \cite{zhu2019dtcdr} designs an adaptable embedding-sharing strategy to combine and share the embeddings of common users across domains. DDTCDR\cite{li2020ddtcdr} develops a latent orthogonal mapping to extract user preferences and maintain inter-domain user similarities.GA-DTCDR \cite{2020A} constructs two separate heterogeneous graphs to generate more representative user and item embeddings and applys an element-wise attention mechanism to blend common user embeddings learned from both domains. PPGN\cite{zhao2019cross} introduces multiple stacked GNN layers to model high order user-item relationships on the interaction graph. BITGCF\cite{liu2020cross} uses graph collaborative filtering network as the base model and proposes a knowledge transfer to aggregate information.However, these methods often fail to effectively distinguish common features and specific features, which would bring the risk of negative transfer. 
\subsection{Disentangled Cross-Domain Recommendation}
Recently, disentanglement-based CDR approaches have received attention.These methods consider domain-invariant and domain-specific features when transfering knowledge across domains.For instance, ATLRec\cite{li2020atlrec} proposes an adversarial transfer learning method to captures domain-invariant features and fuse both domain-specific and domain-invariant features by a linear combination. DisenCDR\cite{cao2022disencdr} proposes the exclusive regularizer and information regularizer to learn meaningful domain-invariant and domain-specific user representations.DCCDR \cite{zhang2023disentangled} employs latent space projection to generate separate user representations and proposes a contrastive learning objective to distinguishable domain-invariant and domain-specific properties to reduce noise and redundancy. DDCDR \cite{DBLP:conf/kdd/AnGYTWHZGG24} uses a teacher-student approach to disentangle and selectively transfer information. GDCCDR\cite{liu2024graph} leverages two distinct contrastive learning constraints for feature disentanglement and employes meta-networks for personalized transfer. While existing methods mainly focus on domain-invariant features, which is intersection preference from both domains. 
Compared with these methoods, our methodology leverages domain-invariant features and integrates two sophisticated mechanisms to effectively capture additional latent valuable non-aligned features, resulting in improved performance.

\section{PROPOSED METHOD}

\begin{figure*}[htbp]
\centering
\includegraphics[width=\textwidth]{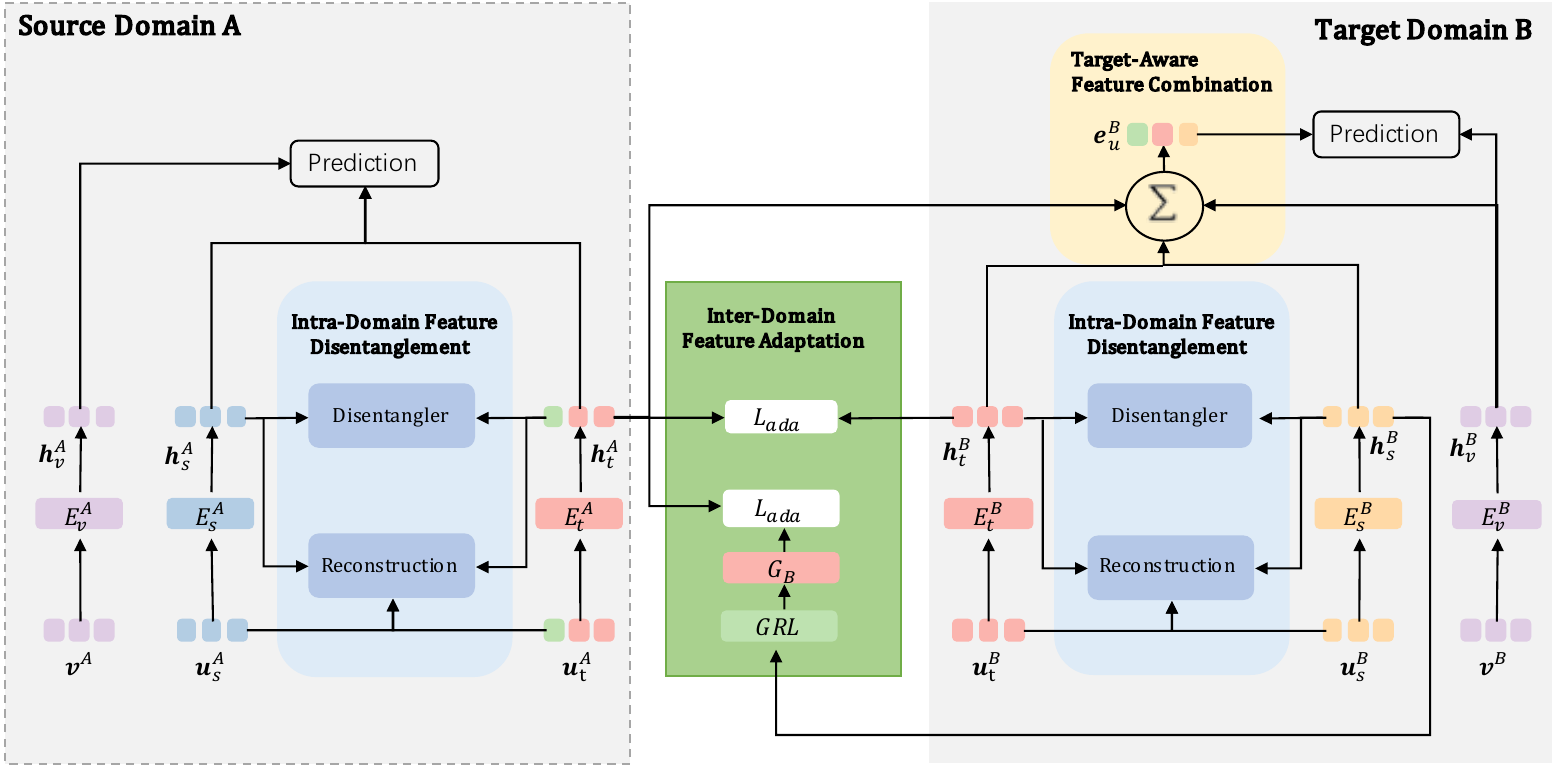} 
\caption{The $A^2DCDR$ architecture has three main modules: Inter-Domain Adaptation, Intra-Domain Disentangler, and Fusion of Representations.
In the Inter-Domain Adaptation module,  the model aligns cross-domain invariant representations and uses adversarial learning to enhance source domain representations.
Next, in the Intra-Domain Disentangler module, a mutual information minimizer is used to separate the domain-specific representation, while a feature reconstructor preserves information integrity during disentanglement.
Finally, in the Fusion of Representations module,  the model combines domain-invariant and valuable non-aligned representations with original contextual features to generate the final prediction.}
\label{fig:framework}
\end{figure*}

In this section, we present the proposed $A^2DCDR$ model, which is depicted in Fig. \ref{fig:framework}. The proposed method consists of three main components:

\begin{itemize}
    \item \textbf{Inter-Domain Feature Adaptation}: 
    We introduce a novel Domain-Constrained Maximum Mean Discrepancy (DC-MMD) technique to effectively capture domain-encompassing features from the source domain while distinguishing them from domain-specific features. This process is crucial for capturing the transferable information that spans across different domains while also preserving the unique characteristics inherent to the target domain.
    
    \item \textbf{Intra-Domain Feature Disentanglement}: 
    To further refine the disentanglement within the same domain, we minimize the mutual information between the domain-encompassing and domain-specific representations. This ensures that the representations remain independent and disentangled, avoiding the overlap of information within the same domain and thereby enhancing the clarity and utility of the features. Moreover, to mitigate information loss during the disentanglement process, we introduce a feature reconstructor to recover the original features from the disentangled representations.

    \item \textbf{Target-Aware Feature Combination}:Finally, we perform adaptive fusion on the disentangled representations over candidates. This step effectively distinguishes the contribution of each type of representation, allowing the model to leverage both transferable and unique information for cross-domain recommendations.
\end{itemize}

We will detail each component as follows.

\subsection{Problem Definition and Notations}
Let $\mathcal{D}^A=(\mathcal{U},\mathcal{V}^A,\mathcal{R}^A)$ and $\mathcal{D}^B=(\mathcal{U},\mathcal{V}^B,\mathcal{R}^B)$ denote two distinct domains which share the same set of users denoted as $\mathcal{U}$. $\mathcal{V}^A$ and $\mathcal{V}^B$ denotes non-overlapped item set in each domain. $\mathcal{R}^A$ and $\mathcal{R}^B$ are two binary interaction matrices $\mathcal{R}^A\in\{0,1\}^{|\mathcal{U}|\times|\mathcal{V}^A|}$ and $\mathcal{R}^B\in\{0,1\}^{|\mathcal{U}|\times|\mathcal{V}^B|}$ for $\mathcal{D}^A$ and $\mathcal{D}^B$ respectively. $\mathcal{R}_{ij}=1$ indicates that user $i$ has interacted with item $j$.

\subsection{Inter-Domain Feature Adaptation}

In response to the challenge of data sparsity and the need for improved recommendation accuracy across different domains, we introduce a novel framework that effectively integrates cross-domain alignment with the exploitation of non-aligned features. Our approach utilizes adversarial training to enhance the transfer of domain-invariant representations, while simultaneously capturing valuable domain-specific characteristics. This dual strategy facilitates comprehensive feature disentanglement, thereby significantly boosting the effectiveness of cross-domain recommendations.

\subsubsection{\textbf{Encoding Separate Representations}}

To embed users and items into low-dimensional dense vectors, we use six \textit{d}-dimensional embedding matrices. For users, we employ $\mathcal{U}_t^A$, $\mathcal{U}_s^A$, $\mathcal{U}_t^B$, and $\mathcal{U}_s^B$, each in $\mathbb{R}^{|\mathcal{U}| \times \textit{d}}$. For items, we have $\mathcal{V}^A \in \mathbb{R}^{|\mathcal{V}^A| \times \textit{d}}$ and $\mathcal{V}^B \in \mathbb{R}^{|\mathcal{V}^B| \times \textit{d}}$. These matrices represent users and items in a shared low-dimensional space, facilitating cross-domain interactions.

In domain $\mathcal{D}^A$, a user $i$ is characterized by a domain-encompassing embedding $\mathbf{u}^A_{ti} = \mathcal{U}_t^A[i,:]$ and a domain-specific embedding $\mathbf{u}^A_{si} = \mathcal{U}_s^A[i,:]$. Similarly, in domain $\mathcal{D}^B$, user $i$ is represented by $\mathbf{u}^B_{ti}$ and $\mathbf{u}^B_{si}$. For item $j$, the embedding in domain $\mathcal{D}^A$ is $\mathbf{v}^A_j = \mathcal{V}^A[j,:]$, while in domain $\mathcal{D}^B$, it is $\mathbf{v}^B_j$. This construction clearly differentiates between the domain-encompassing and domain-specific components of both user and item embeddings. Such a distinction is crucial for facilitating effective knowledge transfer across domains while preserving the unique characteristics inherent to each domain. By constructing these embeddings prior to encoding, our approach precludes conflicts at the most fundamental level of representation, thereby enhancing the model's ability to capture and utilize nuanced cross-domain information.

To enrich the representations, we utilize encoders to capture high-order collaborative information \cite{zhang2023disentangled}. Specifically, in domain $\mathcal{D}^A$, we define three encoders: $E^A_t$, $E^A_s$, and $E^A_v$. $E^A_t$ and $E^A_s$ enrich the domain-encompassing and domain-specific representations, respectively, while $E^A_v$ is  captures high-order item representations.





These encoders can be implemented on top of various network architectures, such as Graph Neural Networks (GNNs)\cite{cao2022disencdr,zhang2023disentangled}, Fully Connected Networks (FNNs), and Transformers\cite{vaswani2017attention}. Inspired by \cite{zhang2023disentangled}, we employ LightGCN\cite{he2020lightgcn} as the encoder for the public dataset. However, considering the efficiency requirements in industrial scenarios, we utilize Transformers as encoders in our industrial scenarios. The specific architecture of these networks, including the number and size of layers, is fine-tuned via cross-validation to achieve optimal performance.

Mathematically, the enrichment representations are obtained as follows:
\begin{align}
    \mathbf{h}^A_t &= E^A_t(\mathbf{u}^A_t) \\
    \mathbf{h}^A_s &= E^A_s(\mathbf{u}^A_s) \\
    \mathbf{h}^A_v &= E^A_v(\mathbf{v}^A)
\end{align}

The enrichment process is identical for domain $\mathcal{D}^B$. 
The raw embeddings generated by the encoders require alignment for effective feature adaptation. To facilitate knowledge transfer, we align the domain-invariant features representations across different domains while preserving valuable non-aligned features. This necessitates a specialized domain adaptation technique, detailed in the following subsection.

\subsubsection{\textbf{Domain-Constrained MMD}}

Existing cross-domain recommendation methods, like DisenCDR and DCCDR, often emphasize domain-invariant features assumed to come from a uniform distribution, which can result in suboptimal recommendations. While these methods align invariant features across domains, they frequently neglect valuable non-aligned features. To address this, we propose the Domain-Constrained Maximum Mean Discrepancy (DC-MMD) regularization technique. 
 


DC-MMD is built upon the concept of Maximum Mean Discrepancy (MMD)\cite{2014Deep}, a statistical measure used to compare the distributions of two sets of data. MMD measures the distance between the mean embeddings of the two distributions in a reproducing kernel Hilbert space (RKHS). It is widely used in domain adaptation to align feature distributions between source and target domains.

In DC-MMD, two key regularizations are applied to facilitate domain adaptation. Firstly, a cross-domain MMD regularization is applied to $\mathbf{h}^A_t$ and $\mathbf{h}^B_t$ to align domain-invariant representations between domains $\mathcal{D}^A$ and $\mathcal{D}^B$. This alignment ensures that the learned representations are effective across both domains. Secondly, another cross-domain MMD regularization is applied in conjunction with a Gradient Reversal Layer (GRL) to act on both $\mathbf{h}^A_t$ and $\mathbf{h}^B_s$. Inverts gradients ($-\nabla$) to force $\mathbf{h}^A_t$ away from $\mathbf{h}^B_s$. This approach allows $\mathbf{h}^B_s$ to learn domain-specific features while simultaneously encouraging $\mathbf{h}^A_t$ to implicitly acquire more non-aligned but valuable information from domain A(e.g., Sport $\rightarrow$ Cloth: ‘gym frequency’). As a result, $\mathbf{h}^A_t$ develops domain-encompassing features, effectively capturing a comprehensive range of characteristics across the domains.

The regularization is formulated as follows:
\begin{equation}
\begin{aligned}
  L_{DC-MMD} = ||\frac{1}{N}\sum\limits_{i=1}^{N}\phi(\mathbf{h}^A_{t,i})-\frac{1}{N}\sum\limits_{i=1}^{N}\phi(\mathbf{h}^B_{t,i})||_{\mathcal{H_\kappa}} \\
  + ||\frac{1}{N}\sum\limits_{i=1}^{N}\phi(\mathbf{h}^A_{t,i})-\frac{1}{N}\sum\limits_{i=1}^{N}\phi(G_B(GRL(\mathbf{h}^B_{s,i})))||_{\mathcal{H_\kappa}}
  \end{aligned}
\end{equation}
where $N$ is the samples numbers that contains $N$ pairs of $(\mathbf{h}^A_t, \mathbf{h}^B_t)$ and $(\mathbf{h}^A_t, \mathbf{h}^B_s)$, where the $i^{th}$ data points is denoted by $\mathbf{h}^A_{t,i}$,$\mathbf{h}^B_{t,i}$ and $\mathbf{h}^B_{s,i}$. $\mathcal{H_\kappa}$ is the reproducing kernel Hilbert space(RKHS) endowed with a characteristic kernel $\kappa$. $\phi$ is the kernel function. $G_B$ is a two-layer neural network. The first term transfer domain-invariant knowledge by aligning feature distributions and the second term aims to remove the domain-invariant information from  $\mathbf{h}^A_s$, thereby achieving effective feature disentanglement between $\mathbf{h}^B_t$ and $\mathbf{h}^A_s$.

DC-MMD captures both domain-invariant and non-aligned features, creating a encompassing representation. This enhances performance in personalized content recommendation and cross-domain advertising by ensuring consistency across domains while leveraging unique domain-specific information.

\subsection{Intra-Domain Feature Disentanglement}

To achieve deep domain-specific embeddings, relying solely on inter-domain feature disentanglement is insufficient, as it indirectly disentangles both $\mathbf{h}^A_t$ and $\mathbf{h}^A_s$, as well as $\mathbf{h}^B_t$ and $\mathbf{h}^B_s$. Moreover, while $\mathbf{h}^A_t$ and $\mathbf{h}^B_t$ may exhibit similar data distributions, this does not imply that they contain identical information. Therefore, to further enhance effective disentanglement, we introduce intra-domain feature disentanglement.

\subsubsection{MI-Based Disentanglement}

Intra-domain feature disentanglement aims to minimize the mutual information (MI) between domain-encompassing and domain-specific representations ($\mathbf{h}^A_t$, $\mathbf{h}^A_s$). MI measures the mutual dependence between two variables from an information-theoretic perspective. For two discrete random variables $\textit{X} \in \mathcal{X}$ and $\textit{Y} \in \mathcal{Y}$, MI is defined as:
\begin{equation}
  \textit{I}(\textit{X}, \textit{Y}) = \int_{\mathcal{X}} \int_{\mathcal{Y}} p(x, y) \log \left( \frac{p(x, y)}{p(x) p(y)} \right) \, dx \, dy,
\end{equation}
where $p(x, y)$ is the joint distribution, and $p(x)$ and $p(y)$ are the marginal distributions. 

However, the closed form of the joint distribution is typically unknown, making MI difficult to compute exactly. 
To address this challenge, we leverage the Contrastive Log-ratio Upper Bound (CLUB) method, introduced by \cite{2020CLUB}, to estimate MI using deep neural networks. Compared to methods that try to optimize the lower limit of mutual information, such as InfoNCE\cite{oord2018representation} and MINE\cite{belghazi2018mutual}, CLUB\cite{2020CLUB} can effectively optimize the upper limit of mutual information, demonstrating superior advantages in the disentanglement of information\cite{xia2023achieving}.

Unlike traditional MI estimation methods, CLUB provides an upper bound that is particularly suitable for our objective of MI minimization.


After convergence, the MI loss for domain $\mathcal{D}^A$ is formulated as:

\begin{equation}
  L^A_{mi} = \frac{1}{N} \sum_{i=1}^{N} \left[ \log q_{\theta}(\mathbf{h}^A_{t,i} | \mathbf{h}^A_{s,i}) - q_{\theta}(\tilde{\mathbf{h}^A}_{t,i} | \mathbf{h}^A_{s,i}) \right],
\end{equation}
where $\tilde{\mathbf{h}^A}_{t,i}$ denotes the $i^{th}$ domain-encompassing representation after shuffling in the samples. $\{(\mathbf{h}^A_{t,i}, \mathbf{h}^A_{s,i})\}_{i=1}^{N}$ represents the $i^{th}$ pair of data points. 
Similarly, $L^B_{mi}$ can be used to encourage disentanglement between $\mathbf{h}^B_t$ and $\mathbf{h}^B_s$ in domain $\mathcal{D}^B$. The total MI loss is defined as: 
\begin{equation}
  L_{mi} = \beta_A L^A_{mi} + \beta_B L^B_{mi},
\end{equation}
where $\beta_A$ and $\beta_B$ are hyperparameters.

By applying CLUB in this context, we provide a practical and efficient solution to MI estimation for domain-specific disentanglement. This approach not only ensures effective disentanglement but also facilitates better cross-domain recommendations by preserving essential information while minimizing redundancy.

\subsubsection{Feature Reconstruction for Information Integrity}

Maintaining information integrity during the disentanglement process is crucial, as the feature disentangler alone may not suffice to preserve all relevant information \cite{peng2019domain}. However, existing DCDR methods \cite{cao2022disencdr,zhang2023disentangled} have overlooked this issue. To mitigate potential information loss, we introduce feature reconstructors $R_A$ and $R_B$, designed to recover the original features from the disentangled representations in both domains. The input of $R_A$ is the concatenation of $\mathbf{h}^A_t$ and $\mathbf{h}^A_s$.

Specifically, let $\widehat{\mathbf{u}^A}$ and $\widehat{\mathbf{u}^B}$ represent the reconstructed features for domains $\mathcal{D}^A$ and $\mathcal{D}^B$, respectively. The reconstruction loss aims to minimize the discrepancy between the original features $\mathbf{u}^A$, $\mathbf{u}^B$ and their reconstructed counterparts $\widehat{\mathbf{u}^A}$, $\widehat{\mathbf{u}^B}$. $\mathbf{u}^A$ is formed by concatenating $\mathbf{u}^A_t$ and $\mathbf{u}^A_s$. Formally, the reconstruction loss $L_{rec}$ is defined as:

\begin{equation}
    L_{rec} = \frac{1}{N} \sum_{i=1}^{N} \left( \gamma_A\| \widehat{\mathbf{u}^A_i} - \mathbf{u}^A_i \|^2_F + \gamma_B\| \widehat{\mathbf{u}^B_i} - \mathbf{u}^B_i \|^2_F \right),
\end{equation}
where $\| \cdot \|^2_F$ denotes the Frobenius norm, effective for measuring the difference between matrices, $\gamma_A$ and $\gamma_B$ are hyperparameters. This loss function ensures that the reconstructed features $\widehat{\mathbf{u}^A}$ and $\widehat{\mathbf{u}^B}$ are closely match the original features $\mathbf{u}^A$ and $\mathbf{u}^B$, thereby preserving the essential information during the disentanglement process.

Integrating the reconstruction loss into our framework is essential for preserving the fidelity of the disentangled representations, ensuring they retain the necessary information for accurate cross-domain recommendations.

\subsection{Target-Aware Feature Combination} 

Existing CDR methods adopt a concatenation operation\cite{zhang2023disentangled} or weighted fusion\cite{2020A} to fuse the representations, while it still struggles in how to effectively distinguish the contribution of each representation. Howerver, according to \cite{zhou2018deep}, the recommendation might be sub-optimal if we directly match the static user features with candidate item set. To this end, we adopt a Target-Aware Feature Combination (TAFC) mechanism to fuse disentangled representations over candidates. 
Moreover, unlike other methods that only use domain-invariant and domain-specific information from the target domain, our approach also incorporates a comprehensive representation from the source domain using DC-MMD, enhancing cross-domain recommendation effectiveness. Formally, the ultimate user $i$ representation for item $j$ in $\mathcal{D}^A$ can be formulated as :
\begin{equation}
\begin{split}
    \mathbf{e}^A_{u_{i,j}} = \sum\limits_{\mathbf{h}_i^A\in{\{\mathbf{h}^{B,t}_i,\mathbf{h}^{A,t}_i,\mathbf{h}^{A,s}_i\}}}{Attn(\mathbf{h}^A_{v_j}, \mathbf{h}_i^A,\mathbf{h}_i^A)}, \\
    Attn(\mathbf{h}^A_{v_j}, \mathbf{h}_i^A, \mathbf{h}_i^A) = Softmax(\frac{<\mathbf{h}^A_{v_j}, \mathbf{h}_i^A>}{\sqrt{d}})\mathbf{h}_i^A.
\end{split}
\end{equation}
where $<.,.>$ represents the dot product.Finally, we utilize the dot product to obtain the final predicted value $\hat{r}^A_{ui_A}$ between user $i$ and item $j$ in domain $\mathcal{D}^A$:
\begin{equation}
    \hat{r}^A_{u_{i,j}} = <\mathbf{e}^A_{u_{i,j}},\mathbf{h}^A_{v_j}>
\end{equation}
Note that $\hat{r}^B_{u_{i,j}}$ can be achieved through a similar process.We use the cross-entropy loss function as follows:
\begin{equation}
         L^A_{ce} = \frac{1}{B} \sum_{i=1}^{B}(r^A_ilog\hat{r}^A_i+(1-r^A_i)log(1-\hat{r}^A_i)),\\
\end{equation}
where $r^A_i$ is $i^{th}$ sample label in a mini batch $B$. We define $L_{ce} = L^A_{ce} + L^B_{ce}$.

The overall objective functions for training of $A^2DCDR$ can be summarized as follows:
\begin{equation}
    L_{total} = L_{ce} + \alpha{L_{DC-MMD}} + {L_{mi}} + {L_{rec}}
\end{equation}
where $\alpha$ is hyperparameter.

\section{EXPERIMENTS}

To evaluate our proposed $A^2DCDR$, we conduct extensive experiments on both public and large-scale industrial datasets. We will answer the following research questions in this section,
\begin{itemize}
\item \textbf{RQ1}: How does $A^2DCDR$ outperform other representative and state-of-the-art methods?
\item \textbf{RQ2}: How do the designed key components of our model contribute to performance improvement?
\item \textbf{RQ3}: How does the robustness of our method at various levels of sparsity in user interaction data? 
\item \textbf{RQ4}: Do the learned representations achieve the expected outcomes?
\end{itemize}

\subsection{Dataset and Evaluation Metrics}

\begin{table}[htbp]
\caption{Experimental Datasets}
\label{tab:Datasets}
\centering
\begin{tabular}{c|c|c|c|c|c}
\hline
Domain & Users & Items & Training & Test & Density \\ \hline
Phone  & 3325    & 38706   & 115554   & 2560 & 0.091\% \\ 
Elec   & 3325    & 17709   & 50407    & 2559 & 0.089\% \\ \hline
Sport  & 9928    & 30796   & 92612    & 8326 & 0.033\% \\ 
Cloth  & 9928    & 39008   & 87829    & 7540 & 0.024\% \\ \hline
Sport  & 4998    & 20845   & 50558    & 3698 & 0.052\% \\ 
Phone  & 4998    & 13655   & 42446    & 3999 & 0.068\% \\ \hline
Elec   & 15761   & 51447   & 210865   & 13824 & 0.027\% \\ 
Cloth  & 15761   & 48781   & 121083   & 12526 & 0.017\% \\ \hline
\end{tabular}
\end{table}

\textbf{Datasets}.We conduct extensive offline evaluations on four real-world datasets from Amazon and conducted an online A/B test in our company to evaluate the performance of our proposed model. Following\cite{cao2022disencdr}, we select Phone, Elec, Cloth and Sport and combine them into four CDR scenarios:Elec\&Phone, Sport\&Cloth, Sport\&Phone and Elec\&Cloth. We use the same training and test set with DisenCDR.The detailed statistics of the datasets are listed in Table \ref{tab:Datasets}.

\textbf{Evaluation Metrics}.Following \cite{cao2022disencdr,zhang2023disentangled}, $A^2DCDR$ predicts 1000 candidates score consisting of 999 random negative items and 1 positive item. Two widely-used metrics HR (Hit Ratio) and NDCG (Normalized Discounted Cumulative Gain) are used to evaluate performance on the top-10 ranking result.

\subsection{Baseline Models and Experimental Setup}
\textbf{Compared Methods.} To validate the effectiveness of $A^2DCDR$, we comprehensively compare it with state-of-the-art (SOTA) single-domain and cross-domain recommendation models. Our benchmark design follows three principles: (1) \textit{Technical evolution} spanning matrix factorization to graph neural networks (GNNs) and disentangled representation learning; (2) \textit{Methodological diversity} covering alignment strategies, graph propagation, variational inference, and contrastive learning; (3) \textit{Temporal validity} including 2023-2024 SOTA methods. To ensure fair comparison, we re-implement baselines using their official code with identical training epochs and early stopping criteria. 

\textbf{Single-Domain Approaches:}
\begin{itemize}
    \item \texttt{NeuMF}~\cite{he2017neural}: Integrates neural networks with matrix factorization to capture non-linear user-item interactions, representing early neural recommendation architectures.
    \item \texttt{NGCF}~\cite{wang2019neural}: Leverages stacked GNN layers to propagate collaborative signals through high-order connectivity.
    \item \texttt{LightGCN}~\cite{he2020lightgcn}: A streamlined GNN variant that removes feature transformations and nonlinear activations, widely recognized as the SOTA single-domain GNN baseline. 
\end{itemize}

\textbf{Cross-Domain Approaches:}
\begin{itemize}
    \item \texttt{DDTCDR}~\cite{li2020ddtcdr}: Aligns user representations across domains via orthogonal mapping while preserving user similarity.
    \item \texttt{PPGN}~\cite{zhao2019cross}: Constructs cross-domain preference graphs to model high-order user-item interactions.
    \item \texttt{BiTGCF}~\cite{liu2020cross}: Facilitates bidirectional knowledge transfer through graph collaborative filtering.
    \item \texttt{DisenCDR}~\cite{cao2022disencdr}: Disentangles domain-shared and domain-specific user preferences using variational inference.
    \item \texttt{DCCDR}~\cite{zhang2023disentangled}: Enhances disentangled representation learning via contrastive objectives.
    \item \texttt{DDCDR}~\cite{DBLP:conf/kdd/AnGYTWHZGG24}: Employs teacher-student networks for selective information transfer, recognized as the 2024 SOTA.
    \item \texttt{GDCCDR}~\cite{liu2024graph}: Leverages two distinct contrastive learning constraints for feature disentanglement, while meta-networks enable the personalized transfer of domain-invariant features, achieving SOTA performance in 2024.

\end{itemize}

\textbf{Experimental Setup}
As used in \cite{cao2022disencdr}, the side information of users/items is not exploited in our experiments. The dimension of both the user and item embeddings is set to 128 for all methods. The learning rate is is selected from \{0.00175, 0.0018, 0.00185, 0.0019, 0.00195, 0.002\}. and optimizer is Adam\cite{kingma2014adam}.The batch size is set to 1024.Besides, each encoder is a 2-layer LightGCN network. Feature reconstructor $R$ is a 1-layer(256) network and $G_B$ is a 2-layer(64,128) network. The activation function of $R$ and $G_B$ is ReLU. The $\alpha$ is 1.0, the $\beta_A$ is 1e-4, the $\beta_B$ is 9e-4,  the $\gamma_A$ is 0.01 and the $\gamma_B$ is 0.09 respectively. These values were determined through a grid search with the following ranges: $\beta \in$ [1e-5, 1e-3]) (step size of 1e-5), $\gamma \in$ [0, 0.1]) (step size of 0.01), and $\alpha \in$ [0, 2]) (step size of 0.1). We train all models with 100 epochs for convergence and run two times for each CDR scenario to botain both domains evaluation results. All experiments are conducted five times with different random seeds, and the average value is provided.

\subsection{Performance Comparisons(RQ1)}
Table \ref{tab:performance_comparison}  lists comparison results on the four CDR scenarios using HR@10 and NDCG@10. we have the following observations.  
First, CDR methods generally outperform single-domain methods, indicating that designed transfer strategies effectively mitigate data sparsity. 
Second, GDCCDR, DDCDR, DCCDR and DisenCDR achieve significant improvements over other methods, demonstrating that learning disentangled representations and transferring domain-invariant representations can further enhance model performance. Third, among all the compared methods, our proposed $A^2DCDR$ achieves the best performance across four datasets. It suggests that our method performes well in efficiently feature disentanglement, information integrity and feature combination. $A^2DCDR$ outperforms full-alignment baselines such as GDCCDR, DDCDR, DCCDR, and DisenCDR, highlighting the contribution of non-aligned features. It is worth noting that the training time complexity is acceptable. On public datasets (e.g., Sport $\rightarrow$ Cloth), our model trains in 30 seconds per epoch on a single NVIDIA P40 GPU, compared to 18 seconds for DDCDR and 88 seconds for DisenCDR, while achieving a +6.43 improvement in HR@10 over DDCDR.
Forth, for Sport\&Phone dataset which contains the fewest training samples,   $A^2DCDR$ achieves the most significant improvement compared to other results. This indicates that $A^2DCDR$ performs better in scenarios with data sparsity.

\begin{table*}[htbp]
    \centering
    \caption{Performance comparison (\%) of different methods on four CDR scenarios.}
    \label{tab:performance_comparison}
    \resizebox{\textwidth}{!}{
    \begin{tabular}{c|c|ccc|ccccccc|c}
        \toprule
        \multirow{2}{*}{\textbf{Datasets}} & \multirow{2}{*}{\textbf{Metrics@10}} & \multicolumn{3}{c|}{\textbf{Single-Domain Methods}} & \multicolumn{7}{c|}{\textbf{Cross-Domain Methods}} & \multirow{2}{*}{\textbf{Ours}} \\
         \cmidrule{3-6} \cmidrule{7-12} 
        & & \textbf{NeuMF} & \textbf{NGCF} & \textbf{LightGCN} & \textbf{DDTCDR} & \textbf{PPGN} & \textbf{BiTGCF} & \textbf{DCCDR} & \textbf{DisenCDR} & \textbf{DDCDR} & \textbf{GDCCDR} \\
        \midrule
        \multirow{2}{*}{Elec} & HR  & 16.17 & 18.55 & 19.17 & 18.47 & 21.68 & 22.14 & 22.58 & 24.57 & \underline{24.78} & 23.71 & \textbf{25.66* (+0.88)}    \\ 
                              & NDCG  & 9.24 & 10.87 & 10.28 & 11.08 & 11.63 & 12.20 & 14.34 & 14.51 & 14.94 & \underline{15.23} & \textbf{16.30* (+1.07)}  \\ 
        \cmidrule{2-13}
        \multirow{2}{*}{Phone} & HR  & 15.84 & 22.79 & 23.25 & 17.23 & 24.54 & 25.71 & 27.63 & 28.76 & 28.89 &\underline{29.75} & \textbf{30.44* (+0.69)}  \\
                               & NDCG  & 8.02 & 12.38 & 12.72 & 8.58 & 13.34 & 13.93 & 17.86 & 16.13 & 17.93 &\underline{19.10} & \textbf{19.47* (+0.37)}  \\
        \cmidrule{1-13}
        \multirow{2}{*}{Sport} & HR  & 10.74 & 13.13 & 13.19 & 11.86 & 15.10 & 14.83 & 16.84 & 17.55 & 17.86 &\underline{19.54} & \textbf{21.25* (+1.71)}  \\
                               & NDCG & 5.46 & 6.87 & 6.94 & 6.37 & 8.03 & 7.95 & 10.05 & 9.46 & 9.77 &\underline{12.03} & \textbf{14.40* (+2.37)}  \\
        \cmidrule{2-13}
        \multirow{2}{*}{Cloth} & HR  & 11.18 & 13.22 & 13.58 & 12.54 & 14.23 & 14.68 & 17.30 & 16.31 & 18.32 &\underline{20.56} & \textbf{23.73* (+3.17)}  \\
                               & NDCG  & 6.02 & 6.97 & 7.29 & 7.13 & 7.68 & 7.93 & 11.49 & 9.03 & 10.94 &\underline{14.35} & \textbf{16.84* (+2.49)}  \\
        \cmidrule{1-13}
        \multirow{2}{*}{Sport} & HR  & 10.11 & 16.06 & 16.33 & 12.14 & 18.00 & 18.63 & 17.82 & 20.17 & 19.25 & \underline{20.57} & \textbf{24.45* (+3.88)}  \\
                               & NDCG  & 5.19 & 8.53 & 9.16 & 6.47 & 10.54 & 10.11 & 10.97 & 11.80 & \underline{13.42} &12.32 & \textbf{15.90* (+2.48)}  \\
        \cmidrule{2-13}
        \multirow{2}{*}{Phone} & HR  & 14.67 & 17.07 & 16.47 & 16.17 & 20.40 & 21.10 &20.83 & 23.55 & \underline{24.02} &23.17 &\textbf{27.81* (+3.79)}  \\
                               & NDCG  & 7.80 & 9.22 & 8.95 & 8.98 & 11.09 & 11.25 & 12.67 & 12.97 & \underline{13.96} &13.80 & \textbf{18.28* (+5.32)}  \\
        \cmidrule{1-13}
        \multirow{2}{*}{Elec} & HR  & 20.08 & 20.20 & 19.97 & 21.70 & 21.85 & 21.61 & 20.76 & 23.71 & \underline{23.91} & 21.92 &\textbf{24.15* (+0.24)}  \\
                              & NDCG & 11.79 & 11.74 & 10.73 & 13.10 & 12.36 & 12.25 & 13.05 & 13.56 & \underline{14.07} &13.71 & \textbf{14.64* (+0.57)}  \\
        \cmidrule{2-13}
        \multirow{2}{*}{Cloth} & HR & 10.84 & 10.86 & 11.24 & 11.47 & 12.98 & 13.11 & 11.03 & 15.13 & 16.75 & \underline{17.03} &\textbf{23.99* (+6.96)}  \\
                               & NDCG & 5.80 & 5.92 & 6.11 & 6.38 & 6.88 & 6.80 & 6.64 & 8.37 & \underline{10.77} &9.63 & \textbf{14.96* (+4.19)}  \\
        \bottomrule
    \end{tabular}
    }
        * indicates that the improvements are statistically significant for $p < 0.05$ judged with the runner-up result in each case by paired t-test.
    
\end{table*}


\subsection{Abaltion Study(RQ2)}
In this section, we conduct an ablation study to evaluate the importance of each designed module in $A^2DCDR$. Specifically:

\begin{itemize}
    \item \textbf{Inter-domain Only}: This variant employs $LightGCN$ with the proposed inter-domain feature disentanglement. In this setup, the user features in $\mathcal{D}^B$ are obtained by sum pooling $(\mathbf{h}^A_t, \mathbf{h}^B_t, \mathbf{h}^B_s)$.
    \item \textbf{Intra- and Inter-domain}: Building on the \textit{Inter-domain Only} variant, this variant introduces intra-domain feature disentanglement to enhance the separation between $\mathbf{h}^*_t$ and $\mathbf{h}^*_s$.
    \item \textbf{Full Model without TAFC}: Extending the \textit{Intra- and Inter-domain} variant, this variant reconstructs the features obtained by the encoder to ensure no information is lost during decoupling. Compared to $A^2DCDR$, the \textit{Full Model without TAFC} fails to improve the user representations over the candidates due to the replacement of TAFC with sum pooling.
\end{itemize}

Table \ref{tab:ablation study} shows the ablation results on the Sport\&Cloth dataset. We observe significant improvements in both HR@10 and NDCG@10 metrics when any component is applied. \textbf{Inter-domain Only} outperforms \texttt{LightGCN} significantly, highlighting the importance of utilizing DC-MMD for learning domain-encompassing features and feature disentanglement across domains. Notably, it achieves results comparable to \texttt{DisenCDR} and \texttt{DDCDR} by simply adding this module. The improvement observed in \textbf{Intra- and Inter-domain} demonstrates that MI minimization can achieve better feature disentanglement. \textbf{Full Model without TAFC} performs better than \textbf{Intra- and Inter-domain}, indicating that feature reconstruction is beneficial for maintaining information integrity. Finally, \textbf{$A^2DCDR$} shows superior performance compared to \textbf{Full Model without TAFC}, underscoring the importance of TAFC in enhancing user features for better candidate recommendations.

\begin{table}[hbp]
\caption{Results of ablation study}
\label{tab:ablation study}
\begin{tabular}{|c|cc|cl|}
\hline
\multirow{2}{*}{\diagbox{Variants}{Metrics@10}}        & \multicolumn{1}{l|}{HR} & \multicolumn{1}{l|}{NDCG} & \multicolumn{1}{l|}{HR} & NDGG \\ \cline{2-5}
                            & \multicolumn{2}{c|}{Sport}                                 & \multicolumn{2}{c|}{Cloth}           \\ \hline
\textit{LightGCN}                      & \multicolumn{1}{c|}{13.19}  & 6.94                        & \multicolumn{1}{c|}{13.58} & 7.29    \\ \hline
\textit{Inter Only}& \multicolumn{1}{c|}{17.32}  & 9.40                        & \multicolumn{1}{c|}{18.94} & 12.33    \\ \hline
\textit{Intra and Inter}& \multicolumn{1}{c|}{20.30}  & 13.29                       & \multicolumn{1}{c|}{22.10} & 14.88    \\ \hline
\texttt{wo TAFC}& \multicolumn{1}{c|}{21.12}  & 13.75                        & \multicolumn{1}{c|}{22.95} & 16.42    \\ \hline
\textit{$A^2DCDR$} & \multicolumn{1}{c|}{21.25}  & 14.40           & \multicolumn{1}{c|}{23.73} & 16.84         \\ \hline
\end{tabular}
\end{table}

\subsection{Data Sparsity(RQ3)}
To analyze the robustness of our method in addressing data sparsity issues, we conducted further experiments on groups with varying numbers of interactions in the training set. 
From the results in Table \ref{tab:data sparsity}, we derive two fundamental observations: 
(1) Compared to \verb|DisenCDR| and \verb|DCCDR|, our model achieves the best results across almost all groups. This indicates that leveraging domain adaptation, minimizing mutual information and employing target-aware feature combination can achieve efficient disentanglement and  knowledge transfer for both inactive and active users. 
(2) For each Cloth group, $A^2DCDR$ shows significant performance improvement, particularly for inactive users in the Sport domain. This suggests that capture domain-encompassing representations can effectively recognize user preferences even under data sparsity conditions.

\begin{table*}[]
\caption{The impact of data sparsity over different user groups on the Sport\&Cloth dataset.}
\label{tab:data sparsity}
\centering
\setlength{\tabcolsep}{7mm} 
\begin{tabular}{c|c|c|c|c|c|c|c}
\hline
Cloth & Sport & \multicolumn{2}{c|}{DCCDR} & \multicolumn{2}{c|}{DisenCDR} & \multicolumn{2}{c}{Ours} \\ \cline{3-8} 
 &  & HR & NDCG & HR & NDCG & HR & NDCG \\ \hline
\multirow{4}{*}{1-10} & 1-10 & 12.84 & 7.82 & 15.37 & 8.30 & 21.45 & 16.12 \\ \cline{2-8} 
 & 11-20 & 15.29 & 9.79 & 18.51 & 10.62 & 21.65 & 13.98 \\ \cline{2-8} 
 & 21-30 & 14.93 & 9.05 & 19.46 & 10.50 & 23.31 & 14.09 \\ \cline{2-8} 
 & \textgreater 30 & 16.23 & 10.36 & 20.78 & 12.31 & 24.09 & 13.65 \\ \hline
\multirow{4}{*}{11-20} & 1-10 & 13.42 & 8.20 & 15.10 & 8.07 & 19.68 & 14.96 \\ \cline{2-8} 
 & 11-20 & 13.83 & 7.96 & 19.37 & 10.82 & 22.64 & 18.04 \\ \cline{2-8} 
 & 21-30 & 14.47 & 9.15 & 22.37 & 13.87 & 23.63 & 16.67 \\ \cline{2-8} 
 & \textgreater 30 & 10.00 & 5.63 & 15.71 & 8.79 & 18.10 & 12.27 \\ \hline
\multirow{4}{*}{21-30} & 1-10 & 8.47 & 4.63 & 7.91 & 4.75 & 19.04 & 16.10 \\ \cline{2-8} 
 & 11-20 & 15.09 & 10.11 & 11.32 & 4.73 & 21.85 & 13.68 \\ \cline{2-8} 
 & 21-30 & 10.00 & 4.17 & 10.00 & 4.66 & 18.13 & 17.08 \\ \cline{2-8} 
 & \textgreater 30 & 17.07 & 11.00 & 21.95 & 13.52 & 26.85 & 17.01 \\ \hline
\multirow{4}{*}{\textgreater 30} & 1-10 & 8.57 & 5.76 & 10.00 & 5.36 & 23.14 & 18.76 \\ \cline{2-8} 
 & 11-20 & 8.82 & 3.76 & 14.71 & 7.58 & 26.17 & 16.32 \\ \cline{2-8} 
 & 21-30 & 15.38 & 11.54 & 15.38 & 4.54 & 17.92 & 14.94 \\ \cline{2-8} 
 & \textgreater 30 & 16.00 & 12.52 & 32.00 & 22.24 & 34.00 & 25.66 \\ \hline
\end{tabular}
\end{table*}

\subsection{Representation Visualization (RQ4)}
\begin{figure}[htbp]
\centering
\includegraphics[width=\linewidth]{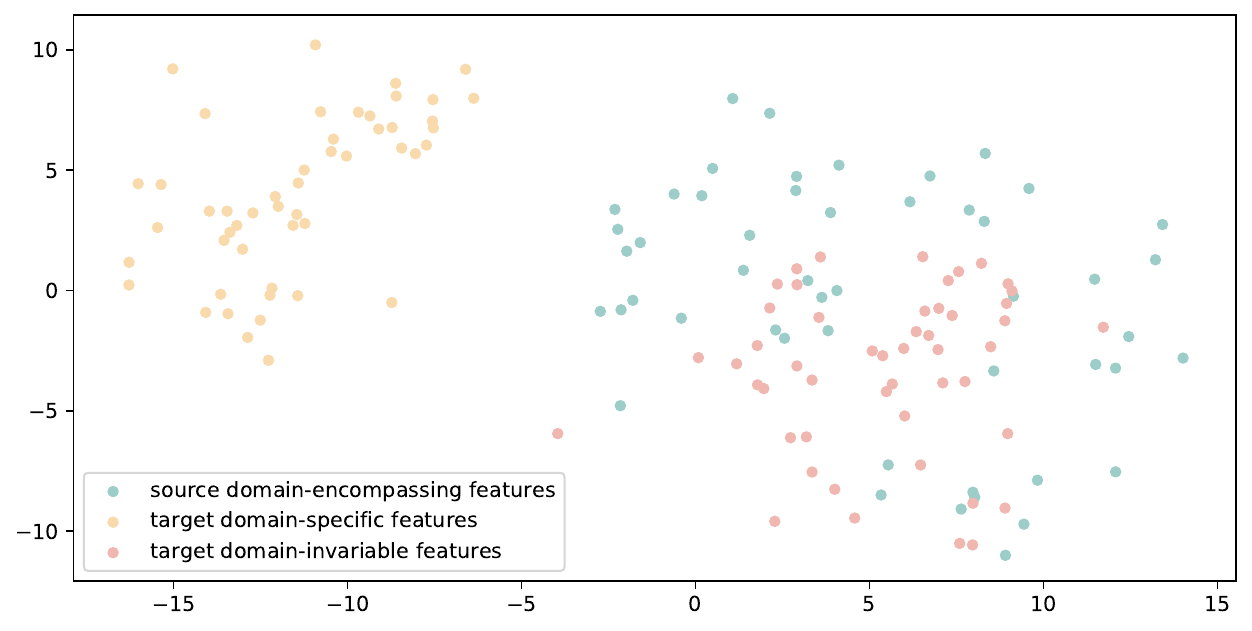}
\caption{t-SNE visualization of disentangled representations in Sport-Cloth domain pair.
Observations: 
(1) Clear separation between the green (source domain-encompassing features, $\mathbf{h}^A_t$) and yellow (target domain-specific features, $\mathbf{h}^B_s$) clusters; 
(2) Overlapping centroids of the red (target domain-invariant features, $\mathbf{h}^B_t$) and green distributions; 
(3) Red features nested within the green cluster boundaries.
}
\label{fig:Visualization}
\end{figure}

To better understand what kind of representations we have learned, we use the t-SNE visualisation technique\cite{van2008visualizing} to visualize three used representations as illustrated in Fig. \ref{fig:Visualization}. We sampled 50 random groups of disentangled features ($\mathbf{h}^A_t$, $\mathbf{h}^B_t$, $\mathbf{h}^B_s$) and visualized them in 2D space using t-SNE. It reveals three fundamental characteristics of the learned representations:
\textbf{Feature Disentanglement}: The source-encompassing features (green) and target-specific features (yellow) form visually distinct clusters with minimal overlap, demonstrating effective separation of domain-specific information.
\textbf{Domain Alignment}: Source $\mathbf{h}^A_t$ and target $\mathbf{h}^B_t$ share nearly identical cluster centers, confirming their distributional consistency across domains.
\textbf{Information Hierarchy}: The target-invariable features (red) are fully contained within the spatial boundaries of source-encompassing features (green), while the latter extend beyond the aligned region. This geometric structure indicates that $\mathbf{h}^A_t$ comprehensively preserves:
\begin{inparaenum}[(i)]
\item cross-domain aligned patterns (overlap with $\mathbf{h}^B_t$),
\item source-exclusive knowledge (non-overlapping regions)
\end{inparaenum} 
which synergistically enhance recommendation performance.

\subsection{Online Deployment}
In our e-commerce platform with dual-feed layout for video, product and other content distribution, we deploy $A^2DCDR$ for an A/B test. The test involves 4.4 million active users with an average of 4.9 videos watched per user. As the A/B test aims to optimize the video ranking model, we set videos as the source domain and products as the target domain. To ensure ecological validity, we keep the feature-engineering pipelines identical for both groups and freeze other recommendation components within the ranking model.

\textbf{Architecture Adaptation.} 
In prior research on public datasets, we employed LightGCN as the backbone within the $A^2DCDR$ framework. However, in our online system, the Behavior Sequence Transformer (BST) \cite{chen2019behavior}  has been deployed as the backbone network. Specifically, BST extracts the video and item interest representations, denoted as $h^V_*$ and $h^I_*$ respectively, from user behavior sequences.
We further expand the $A^2DCDR$ framework by leveraging the representations generated by BST. Through the processes of cross domain alignment and disentanglement applied to $h^V_*$ and $h^I_*$, the $A^2DCDR$ framework demonstrates remarkable extensibility.
This extensibility comes from the inherent ability of $A^2DCDR$ to seamlessly integrate with various backbones. Whether LightGCN or BST, $A^2DCDR$ functions efficiently, reducing framework and adapting to varivariousommendation scenarios.

\textbf{Offline Training.} We adopt an incremental training paradigm, where model parameters are updated daily based on the latest user interaction logs. This increases the complexity of the training pipeline, but is acceptable since offline training is not latency-critical. Importantly, most additional computationally intensive operations—such as cross-domain representation alignment, interest decoupling, and advanced regularization (e.g., DC-MMD, GRL, CLUB)—are performed entirely in the offline stage. The training pipeline leverages a distributed GPU cluster based on NVIDIA H100. This design allows us to integrate rich and heterogeneous cross-domain information (e.g., user behaviors on both videos and products), and to leverage complex modeling techniques, without affecting online inference efficiency. The introduction of cross-domain information inevitably increases the size of the embedding vocabulary, but this is effectively managed in the online stage.

\textbf{Online Performance.}  As presented in Table \ref{tab:A/B Testing}, the Click-Through Rate (CTR) comparison results demonstrate a 7.18\% improvement over fifteen consecutive days using the proposed method. This improvement was statistically significant according to t-tests. To efficiently handle the enlarged embedding vocabulary and ensure scalable, low-latency serving, we explicitly decouple the embedding table (vocabulary) management from the inference computation. The embedding tables are stored and managed on a distributed parameter server (PS), which supports sharding, dynamic admission and eviction of embeddings, and efficient access under high QPS. The inference service itself is deployed on a distributed CPU cluster, which independently pulls the required embeddings from the PS for each request. This separation allows us to scale embedding storage and inference computation independently, and efficiently serve complex cross-domain models. As a result, even with the 3.8GB increased embedding increment (baseline: ~100GB baseline) brought by cross-domain modeling, the additional inference overhead from our method remains below 1ms, and the overall P99 latency is within the strict 33ms limit, even with nearly 1,000 candidates per request. Consequently, the $A^2DCDR$ framework can be effectively deployed in online systems. It has been successfully implemented to handle the traffic for e-commerce short videos, operating within a large-scale CPU cluster.

\setlength{\tabcolsep}{4mm}{
\begin{table}
\caption{Offline and Online A/B testing results.}
  \begin{tabular}{ccc}
    \toprule
    Methods & Offline AUC & Online CTR Gain \\
    \midrule
     BST & 71.5 &- \\
     Inter Only & 72.1 & - \\
     wo TAFC & 72.5 & -\\
     $A^2DCDR$ & 72.7 & +7.18\%\\
    \bottomrule
  \end{tabular}
  \label{tab:A/B Testing}
\end{table}}

\section{Conclusion}
In this paper, we propose a novel CDR model, namely Disentangled Domain Distribution for Cross-Domain Recommendation($A^2DCDR$). Different from previous works which focus on domain-shared information, $A^2DCDR$ utilizes domain adaption and MI to effectively disentangle domain-encompassing and domain-specific representations and achieve much better performance. Moreover, the proposed TAFM fuses the disentangled representations with an adaptive manner and further improve the performance. Both offline and online experimental results demonstrate the effectiveness of $A^2DCDR$ over several state-of-the-art methods. In future work, we will extend $A^2DCDR$ to multi-domain scenarios with hierarchical relationships, investigate long-sequence cross-domain dependency modeling, and explore causal counterfactual reasoning for domain-shift effect estimation to enhance recommendation robustness.


\bibliographystyle{ieeetran}
\bibliography{IEEEabrv,sample-base}

\end{document}